\documentclass[amsfonts,amsmath,letterpaper,11pt,prd,nofootinbib,floatfix,tightenlines,preprintnumbers]{revtex4}
\usepackage{graphicx}
\usepackage{amssymb}
\usepackage{bm}
\usepackage{comment}
\usepackage{color} 

\newcommand{\be}{\begin{equation}}
\newcommand{\ee}{\end{equation}}
\newcommand{\bea}{\begin{eqnarray}}
\newcommand{\eea}{\end{eqnarray}}

\newcommand{\Choose}[2]{{\begin{pmatrix} {#1} \\ {#2} \end{pmatrix}}}

\usepackage{dcolumn}
\newcolumntype{q}{D{.}{.}{1.2}}
\newcolumntype{d}{D{.}{.}{2.5}}
\newcolumntype{s}{D{.}{.}{4.8}}

\begin{document}

\title{Finite-volume matrix elements in multi-boson states}
 
\author{William Detmold} \affiliation{ Center for Theoretical Physics,
  Massachusetts Institute of Technology, Cambridge, MA 02139, USA}
\author{Michael Flynn} \affiliation{ Center for Theoretical
  Physics, Massachusetts Institute of Technology, Cambridge, MA 02139,
  USA}
\date{\today}

\begin{abstract}
  We derive the relations necessary for the extraction of matrix 
  elements of multi-hadron systems from finite-volume QCD calculations. 
  We focus on systems of $n\geq2$ weakly interacting identical particles 
  without spin. These results will be useful in extracting physical 
  quantities from lattice QCD measurements of such matrix elements in many-pion and many-kaon systems.
\end{abstract}

\preprint{MIT-CTP {4622}}

\maketitle

\section{Introduction}

An important goal in nuclear physics is to understand how the presence of a hadronic/nuclear medium modifies the properties of hadrons. Experimentally there are a number of examples where such modifications are observed and are significant in their effects. The EMC effect \cite{Aubert:1983xm,Norton:2003cb}, modifications of the parton distribution functions  of the proton inside a nucleus, is a particularly well studied example where ${\cal O}(10\%)$ effects are observed. Similarly,  Gamow-Teller transitions of nuclei occur at rates that indicate that the axial coupling of the nucleon is modified at an even more significant level in medium-mass nuclei, being as large as a 30\% effect in some cases \cite{Krofcheck:1985fg,Chou:1993zz}. It is natural that such effects arise as a result of the strong dynamics that exist inside the nucleus. However, theoretically such effects are not understood in a compelling, predictive way and it is a contemporary challenge to provide a rigorous description of these effects using methods that are directly connected to the underlying theory of the strong interactions, Quantum Chromodynamics (QCD). This is not purely an academic exercise in understanding the structure of a nucleus;  nuclei are becoming increasingly important as targets in contemporary and planned studies of neutrino properties and in many searches for physics beyond the standard model. The ability of  the Long Baseline Neutrino Facility and other proposed neutrino experiments to determine the neutrino mass hierarchy and extract the CP violating phases in the neutrino mixing matrix is limited by neutrino flux and  energy measurements on nuclear targets \cite{Mosel:2013fxa,Coloma:2013tba}. These, in turn,  are fundamentally limited by the current uncertainties in our knowledge of the axial (and induced pseudoscalar) form factors of  nuclei. In many dark matter direct detection experiments, nuclear recoils are the primary signal mechanism. Expected rates therefore depend not only on the dynamics of the dark sector, but also on the 
amplitudes for interactions of the target nuclei (Ar, Si, Ge, Xe,\ldots) with the current that mediates the connection to the dark sector. For example, for a dark sector that couples to the Standard model via a scalar mediator, the relevant Standard Model input is the nuclear target matrix element of the scalar quark bilinear current, the so-called sigma term of the nucleus \cite{Prezeau:2003sv,Beane:2013kca}.
 Understanding nuclear effects in these classes of experiments at a quantitative level is required to maximise their impact and is thus an important goal for QCD practitioners over the coming decade.

In this work, we  develop the theoretical background necessary for the QCD exploration of external currents in particularly simple multi-hadron systems. 
As the only known method of calculating the properties of hadrons (including nuclei) from first principles is through lattice QCD (LQCD),  it is expected that the requisite understanding will involve lattice calculations. However lattice calculations are performed in Euclidean space and in a finite volume by necessity, which restricts the physical (infinite-volume Minkowski space) information that can be extracted. It is important to understand what information is accessible in 
such calculations and how it can be extracted. In its fully generality, this is a very challenging task 
and to make progress, we will  focus on the limiting case of perturbatively interacting spin-zero sysytems 
in our current analysis.

\section{Multi-boson systems}

Over the last few years, systems of many identical composite bosons have been extensively studied in 
lattice QCD with particular focus on states with the quantum numbers of many like-charged pions.
Following the classic works of Lee, Huang and Yang \cite{Huang:1957im,Lee:1957zzb}, the theoretical understanding of the dependence of the ground state spectrum of these systems on the  finite volume used in numerical calculations was developed in 
Refs. \cite{Beane:2007qr,Detmold:2008gh}.
There, the ground-state energy of $n$ identical bosons of mass $M$ in a cubic box of side length $L$ was determined
using time-ordered perturbation theory, 
with a Hamiltonian of the form
\begin{eqnarray}
H & = & 
\sum_{\bf k}\ h_{\bf k}^\dagger \  h_{\bf k}\ \left(\ {|{\bf k}|^2\over 2 M}\ -\
  {|{\bf k}|^4\over 8 M^3}\ \right)
\nonumber\\
&& + \
{1\over (2!)^2}
\sum_{\bf Q,k,p}\  h_{\bf {Q\over 2}+k}^\dagger  h_{\bf {Q\over 2}-k}^\dagger\
h_{\bf {Q\over 2}+p}\ h_{\bf {Q\over 2}-p}
\ \left(\ {4\pi a\over M}\ +\ {\pi a\over M}\left( a r - {1\over 2 M^2}\right)
  \left(\  |{\bf k}|^2 + |{\bf p}|^2\ \right)\ \right)
\nonumber\\
&& + \
{\eta_3(\mu)\over (3!)^2}\ 
\sum_{\bf Q,k,p,r,s}\  h_{\bf {Q\over 3}+k}^\dagger  h_{\bf {Q\over 3}+p}^\dagger\ h_{\bf {Q\over 3}-k-p}^\dagger\
 h_{\bf {Q\over 3}+r}  h_{\bf {Q\over 3}+s}\ h_{\bf {Q\over 3}-r-s}
\ ,
\label{eq:interaction}
\end{eqnarray}
where the operator $h_{\bf k}$ annihilates a boson with momentum ${\bf k}$
with unit amplitude, and terms are kept that will contribute at the  order 
in the large-volume expansion to which we work. The couplings $a$, $r$ and $\eta_3(\mu)$ correspond
the the two-particle scattering length and effective range, and to the leading momentum-independent 
three particle interaction.\footnote{The three-particle interaction $\eta_3(\mu)$ as defined in the Hamiltonian depends on the regularisation and renormalisation prescription as discussed 
in Ref.~\cite{Beane:2007qr}, but will not contribute at the order we work in this current study.} In particular, the shift in the ground-state energy from $n$ free bosons was determined to be 
\begin{eqnarray}
\label{eq:7}
  \Delta E_0(n,L) &=&
  \frac{4\pi\, a}{M\,L^3}\Choose{n}{2}\Bigg\{
  1 -\left(\frac{a}{\pi\,L}\right){\cal I}
+\left(\frac{a}{\pi\,L}\right)^2\left[{\cal I}^2+(2n-5){\cal J}\right]
\\&&\hspace*{2cm}
- \left(\frac{a}{\pi\,L}\right)^3\Big[{\cal I}^3 + (2 n-7)
  {\cal I}{\cal J} + \left(5 n^2-41 n+63\right){\cal K}\Big]
\nonumber
\Bigg\}
+\Choose{n}{2} \frac{8\pi^2 a^3 r }{M\, L^6}\ 
\nonumber
\\&&\hspace*{0cm}
+\Choose{n}{3} {1\over L^6}\ 
\left[\ 
\eta_3(\mu)\ +\ {64\pi a^4\over M}\left(3\sqrt{3}-4\pi\right)\ \log\left(\mu
  L\right)\ -\ 
{96 a^4\over\pi^2 M} \ {\cal S}
\ \right]
\ \ + \ {\cal O}\left(L^{-7}\right)\,,
\nonumber
\end{eqnarray}
where $\mu$ is a remornalisation scale and 
$$
{\cal I}\ =\  -8.9136329\,, \qquad
{\cal J}\ =\  16.532316\,, \qquad
{\cal K}\  = \  8.4019240\,, \qquad
{\cal S}_{\rm MS}\ = \ -185.12506\,,
$$ 
are geometric constants arising from finite-volume loop contributions \cite{Beane:2007qr,Detmold:2008gh}.
The corresponding expression including ${\cal O}(1/L^7)$ corrections is presented 
in Ref.~\cite{Detmold:2008gh}.

Determinations of the corresponding energy shifts in many-boson systems can be used to determine 
the various interactions in Eq.~(\ref{eq:interaction}) for a given set of systems. To this end,
sophisticated techniques have  been constructed in order to study these complicated systems numerically in QCD\cite{Detmold:2010au,Shi:2011mr,Detmold:2012wc}. Calculations using 
these methods have led to extractions of the $I=2$ two-pion and $I=3$ three-pion interactions 
and of the effects of these systems on other hadronic quantities \cite{Detmold:2008bw,Detmold:2012pi}.
 Using relations between baryons and 
mesons in QCD with $N_c=2$ colours, these results have also enabled a recent 
study of the analogues of nuclei for  $N_c=2$ \cite{Detmold:2014kba}.

From considerations of chiral dynamics, QCD inequalities \cite{Detmold:2014iha}, and from the explicit numerical explorations mentioned above, 
it is apparent that interactions in isospin $I=n$ many-$\pi^+$  systems are repulsive and that there are no 
bound states  for any $n$.
Chiral symmetry guarantees that the strength of the interactions is perturbatively weak so an 
expansion in the couplings $a$, $r$ and $\eta_3(\mu)$ is expected to be reliable provided $n a /L$ remains small, as do similar combinations of the other couplings.
Such systems therefore provide an ideal situation for the application of the methods discussed herein.

\section{Matrix elements of external currents in multi-boson systems}

The time-ordered perturbation theory methods used to derive the energy shifts in Refs.~\cite{Beane:2007qr,Detmold:2008gh} 
order by order in the coupling and large-volume expansion also determine the state vector as an expansion
in couplings (see, for example Ref.~\cite{sakuraiMQM}).
In particular, the $n$ boson state can be expanded as
\begin{equation} 
\label{meexp}
|n \rangle (a, r, \eta_3(\mu))= |n^{(0)} \rangle + \eta |n^{(1)} \rangle
 + \eta^2 |n^{(2)} \rangle
  + \eta^3 |n^{(3)} \rangle
+\ldots,
\end{equation}
where $|n^{(0)} \rangle$ corresponds to the free $n$-particle system and 
subsequent terms are induced by perturbative interactions amongst the particles in the periodic 
volume; in the above expression, $\eta$ is representative of any one of the couplings. Knowing the state vector, it is thus a simple matter to
compute the expectation values of  currents that are of phenomenological interest. To be 
general, we do not assume a particular type of current and consider the form
\begin{equation} \label{currentoperator}
J = \sum_{\bf k} \alpha_1 h^{\dag}_{\bf k}h_{\bf k} +
\sum_{\bf k,Q,p}  \alpha_2 h_{\bf {Q\over 2}+k}^\dagger  h_{\bf {Q\over 2}-k}^\dagger\
h_{\bf {Q\over 2}+p}\ h_{\bf {Q\over 2}-p}\,,
\end{equation}
where $\alpha_1$ and $\alpha_2$ are constants that describe the momentum independent
 one-boson current and the two-boson current, respectively. The 
 particular strengths of the different terms, and the flavour and spin dependence of the interactions may differ for different fundamental currents, but the above form is general up to 
 momentum-dependent and higher-body corrections that are suppressed by additional
 powers of $1/L$ in our results.
For simplicity, we work in the soft  limit where the current injects no momentum into the 
system so that the two-hadron current amounts to a simple reshuffling of the boson 
momenta as indicated. 

The full finite volume matrix elements of $J$ involve the various terms in Eq.~(\ref{meexp}).
The calculation is straightforward (if a little tedious) and the 
reader is referred to Refs.~\cite{Beane:2007qr,Detmold:2008gh} for more details; we will only state the result. The matrix elements of $J$ for systems of $n$ pions up to $O(L^{-5})$ are as follows:

\begin{eqnarray}
\langle n  | J| n \rangle &=&
%
n\alpha_{1} +  \frac{n\alpha_{1}a^2}{\pi^{2}L^{2}}{{n}\choose{2}} \mathcal{J}
+ \frac{\alpha_2}{L^{3}}{{n}\choose{2}} \\
&&
+ \frac{2n\alpha_{1}a^{3}}{\pi^{3}L^{3}}{{n}\choose{2}}\left\{\mathcal{K}{{n}\choose{2}} -\left[\mathcal{I\,J} + 4\mathcal{K} {{n-2}\choose{1}} + \mathcal{K}{{n-2}\choose{2}}\right]\right\}
 -\frac{2\alpha_2 a }{\pi L^{4}}{{n}\choose{2}}\mathcal{I} 
 \nonumber \\
&&+ \frac{n\alpha_{1}a^{4}}{\pi^{4}L^{4}}\Bigg[3\mathcal{I}^{2}\mathcal{J} + \mathcal{L}\left(186-\frac{241n}{2}+\frac{29}{2}n^{2}\right)
+\mathcal{J}^{2}\left(\frac{n^{2}}{4}+\frac{3n}{4}-\frac{7}{2}\right)
\nonumber \\ &&
\hspace*{2.3cm}+\mathcal{I\,K}(4n-14)+{\cal U}(32n-64)+{\cal V}(16n-32)\Bigg]
+{\cal O}(1/L^5)\,.
\nonumber
\end{eqnarray} 
This expression is the primary result of the current work and has been calculated through to the second order
at which the two-boson current contributes so that the consistency of an extraction can 
be checked between orders. The  
additional numerical constants that enter this expression are
$$ 
\mathcal{L} = 6.9458079,\qquad
\mathcal{U}= 85.1269266, \qquad
\mathcal{V}=-64.1765107\,,
$$
and the sums which lead to these values are defined by 
\begin{eqnarray}
\mathcal{L} &=& \sum_{\vec{i}\ne0} \frac{1}{|\vec{i}|^8}\,,\\
\mathcal{U} &=& \sum_{\vec{i},\vec{j} \neq 0} \frac{1}{|\vec i|^{4}|\vec{j}|^{2}\left(|\vec i|^{2}+|\vec j|^{2}+|\vec i+\vec j|^{2}\right)} + \sum_{\vec i,\vec j \neq 0}\frac{1}{|\vec i|^{2}|\vec j|^{2}\left(|\vec i|^{2}+|\vec j|^{2}+|\vec i+j\vec|^{2}\right)^{2}}\,,  \\
\mathcal{V}&=& \sum_{\vec i,\vec j \neq 0}\frac{1}{|\vec i|^{6}\left(|\vec i|^{2}+|\vec j|^{2}+|\vec i+\vec j|^{2}    
\right)} + \sum_{\vec i,\vec j \neq 0}\frac{1}{|\vec i|^{4}\left(|\vec i|^{2}+|\vec j|^{2}+|\vec i+\vec j|^{2}\right)^{2}} \,,
\end{eqnarray}
where $\vec i$ and $\vec j$ are three-tuples with integer valued components. These three- and six-dimensional sums are convergent and can be computed with the use of the Poisson summation formula, yielding the values above.

From the above expression, we see that the finite-volume matrix elements  only 
depend on the one-boson current, $\alpha_1$, at leading order  and at next-to-leading order 
in the large volume perturbative expansion. Dependence on the two-boson current 
coupling, $\alpha_2$, arises at ${\cal O}([\frac{a}{\pi L}]^3)$; for a repulsive interaction 
such weak sensitivity is expected. 
Notice  that neither $r$ or $\eta_3(\mu)$ enter the calculation at ${\cal O}(1/L^4)$ however 
they will contribute at higher orders in $1/L$. Similarly, a three-boson contribution to the current will eventually be relevant.
As with the energy levels in Eq.~(\ref{eq:7}), off-shell effects will lead to additional exponentially suppressed volume dependence $\sim \exp(- M_\pi L)$ where $M_\pi$ is the pion mass which domintes such effects 
as the pion is the lightest hadronic state.

\section{Discussion}

The result presented above provides the expected hadronic behaviour of a multi-boson matrix element 
of a local (at the hadronic scale) operator in a finite volume. It explicitly depends on the 
one-body and two-body couplings of the hadrons to the current and also on the two-body 
interactions between the hadrons (higher body interactions will become relevant for 
sub-leading terms in the volume expansion). Lattice QCD calculations of the corresponding 
matrix elements in systems of $n$ spin-zero bosons can be matched onto these expressions to 
determine the external current interactions in the appropriate hadronic theory once the two-boson
interaction is determined from the shifts in energies of $n$-boson systems in a finite volume. 
Consequently, the results derived herein will be useful
in the analysis of lattice QCD calculations of matrix elements of currents in
weakly-interacting multi-pion states such as those presented in preliminary form in Ref.~\cite{Detmold:2011np}.

Our calculation has focused on the case of identical spin zero bosons with perturbatively weak 
interactions at energies near threshold in the appropriate channels. The 
 inclusion of the effects of angular momentum and spin degrees of freedom, and of more complicated systems
 with coupled channels is left for future study.
Further work is also necessary to understand the behaviour of multi-hadron matrix elements 
 with non-perturbatively strong interactions or when the expansion in $a/L$ breaks down. 
 For two particles, the non-perturbative dependence
of the ground state energy on the spatial extent of a periodic volume has been known
for many years \cite{Luscher:1986pf,Luscher:1990ux} and there has been significant recent 
progress \cite{Briceno:2012rv,Polejaeva:2012ut,Hansen:2014eka} toward achieving the same 
level of understanding for three-particle systems. The effects of finite volume on 
$1\to 2$ particle transitions induced by an external current
have also been understood for simple cases in the 
pioneering work of Lellouch and L\"uscher \cite{Lellouch:2000pv} 
and recently generalised to more complicated 
cases in Refs.~\cite{Detmold:2004qn,Meyer:2012wk,Hansen:2012tf,Briceno:2012yi,Agadjanov:2014kha,Briceno:2014uqa}. 
It seems likely that the approaches used in these analyses could be extended to consideration 
of $2\to2$ current matrix elements and perhaps to the three-particle case. For strongly interacting systems with more than three particles, 
new methods are required to have analytic control over the  interactions of multi-hadron systems and  over the relation between multi-hadron matrix
elements in QCD and in the hadronic  theory.
In the absence of  such advances, the matching between QCD calculations of matrix elements in finite 
volume and those in the hadronic 
effective theory can be implemented through numerical calculations of correlators in the hadronic 
theory in a finite volume for varying input low-energy constants (the analogues of the current couplings 
$\alpha_1$ and $\alpha_2$) until the QCD results are reproduced, 
thereby determining the hadronic couplings.

\acknowledgments{
This work was supported by the US Department of Energy Early Career Research Award DE-SC0010495. 
The authors are grateful to  Z. Davoudi, H.-W. Lin and M. J.~Savage for discussions.}

\bibliography{MHME}

\end{document}